\documentclass[twocolumn,aps,prl,preprintnumbers]{revtex4}
\usepackage{bbm}
\usepackage[latin9]{inputenc}
\usepackage{amsmath,amssymb,amsthm,amsfonts,mathrsfs}
\usepackage{graphicx}
\usepackage{color}
\usepackage{txfonts}
\usepackage[colorlinks=true,citecolor=blue,linkcolor=blue,urlcolor=blue,anchorcolor=blue]{hyperref}%
\hypersetup{colorlinks=true,citecolor=blue,linkcolor=blue,urlcolor=blue}

\makeatother
\begin{document}
\title{Twisted Magnon Frequency Comb and Penrose Superradiance}
\author{Zhenyu Wang$^{1}$}
\author{H. Y. Yuan$^{2}$}
\author{Yunshan Cao$^{1}$}
\author{Peng Yan$^{1}$}
\email[Corresponding author: ]{yan@uestc.edu.cn}
\affiliation{$^{1}$School of Electronic Science and Engineering and State Key Laboratory of Electronic Thin Films and Integrated Devices, University of Electronic Science and Technology of China, Chengdu 610054, China}
\affiliation{$^{2}$Institute for Theoretical Physics, Utrecht University, 3584 CC Utrecht, The Netherlands}

\begin{abstract}
Quantization effects of the nonlinear magnon-vortex interaction in ferromagnetic nanodisks are studied. We show that the circular geometry twists the spin-wave fields with spiral phase dislocations carrying quantized orbital angular momentum (OAM). Meanwhile, the confluence and splitting scattering of twisted magnons off the gyrating vortex core (VC) generates a frequency comb consisting of discrete and equally spaced spectral lines, dubbed as twisted magnon frequency comb (tMFC).
It is found that the mode spacing of the tMFC is equal to the gyration frequency of the VC and the OAM quantum numbers between adjacent spectral lines differ by one. By applying a magnetic field perpendicular to the plane of a thick nanodisk, we observe a magnonic Penrose superradiance inside the cone vortex state, which mimics the amplification of waves scattered from a rotating black hole. It is demonstrated that the higher-order modes of tMFC are significantly amplified while the lower-order ones are trapped within the VC gyrating orbit which manifests as the ergoregion. These results suggest a promising way to generate twisted magnons with large OAM and to drastically improve the flatness of the magnon comb.
\end{abstract}

\maketitle
\emph{Introduction}.---Wave fields carrying orbital angular momentum (OAM) have been investigated widely ranging from photons \cite{Allen2003,Torres2011,Andrews2012,Jones2015,Mair2001,Tamburini2011,Vallone2014,Yang2018}, phonons \cite{Anhauser2012,Hong2015,Marzo2018,Baresch2018,ZhangPRAppl2018}, electrons \cite{Bliokh2007,Uchida2010,Verbeeck2010,McMorran2011,Mafakheri2017,Silenko2017,Lloyd2017} to neutrons \cite{Clark2015,Cappelletti2018} and gluons \cite{Ji2017}. The OAM is generally associated with spatially twisted phase profile $e^{il\phi}$ with $\phi$ the azimuthal angle and $l$ being an integer referred to as the topological charge. Twisted spin wave (magnon) represents such a novel state carrying quantized OAM. Peculiar features of twisted magnon include the robustness of the topological charge against the magnetic damping \cite{Chen2020}, efficiency of manipulating magnetic quasiparticles as a magnetic tweezer \cite{Jiang2020}, etc. Twisted magnons can be generated by applying an electric field through the Aharonov-Casher effect \cite{Jia2019,Jia2021}, a spatially inhomogeneous radio-frequency magnetic field \cite{Chen2020,Jiang2020}, and a magnonic spiral phase plate \cite{Jia201902}. However, an efficient method to generate twisted magnons with high OAM is still desired.

\begin{figure}[!h]
  \centering
\includegraphics[width=0.45\textwidth]{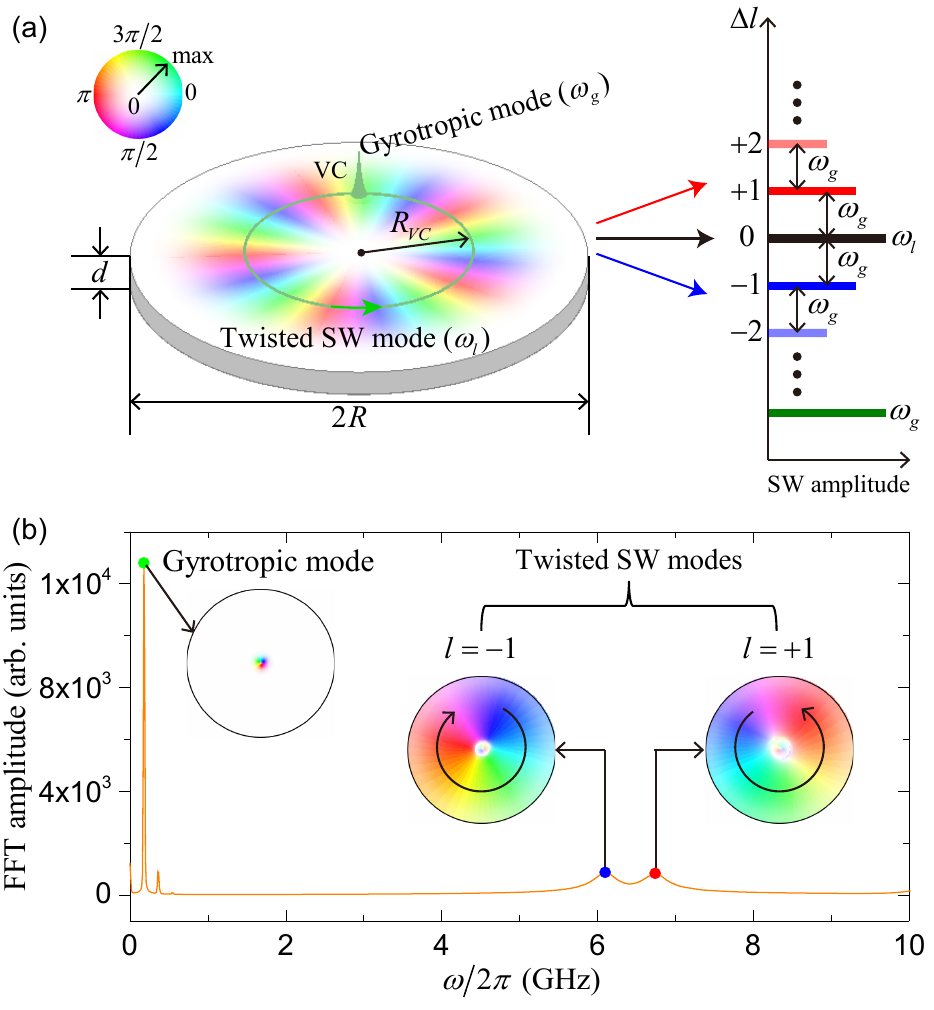}\\
\caption{(a) Schematic illustration of nonlinear interactions between the VC and twisted spin waves (SWs) and the resulting tMFC. The green circle with a arrow indicates the VC gyration direction and trajectory. The inset is the color palette of twisted SWs with color indicating the phase and intensity representing the amplitude. (b) FFT spectrum of the vortex disk. The inset shows the mode profiles of the gyrotropic mode and twisted SW modes $l=\pm1$. Circular curves with arrows label the propagation directions of twisted SWs.}\label{fig1}
\end{figure}

It has been shown that the nonlinear interaction between the planar spin-wave and magnetic texture can generate a frequency comb \cite{Wang2021}. However, the OAM carried by each spectrum line is zero due to the planar nature of the spin wave. Recently, three-magnon splitting in magnetic vortex was observed in experiments, where a radial spin-wave mode splits into two azimuthal spin-wave modes (or twisted magnons) with opposite OAMs \cite{Schultheiss2019}. Besides radial and azimuthal spin-wave modes, there also exists gyrotropic mode of the vortex core (VC). Interactions between the VC and azimuthal spin waves have been reported with focus mainly on the VC-induced frequency splitting of azimuthal spin waves \cite{Park2005,Guslienko2008} and the VC reversal driven by azimuthal spin waves \cite{Kammerer2011,Yoo2015}. The nonlinear interaction between them however is rarely addressed.

\begin{figure}
\centering
  \includegraphics[width=0.49\textwidth]{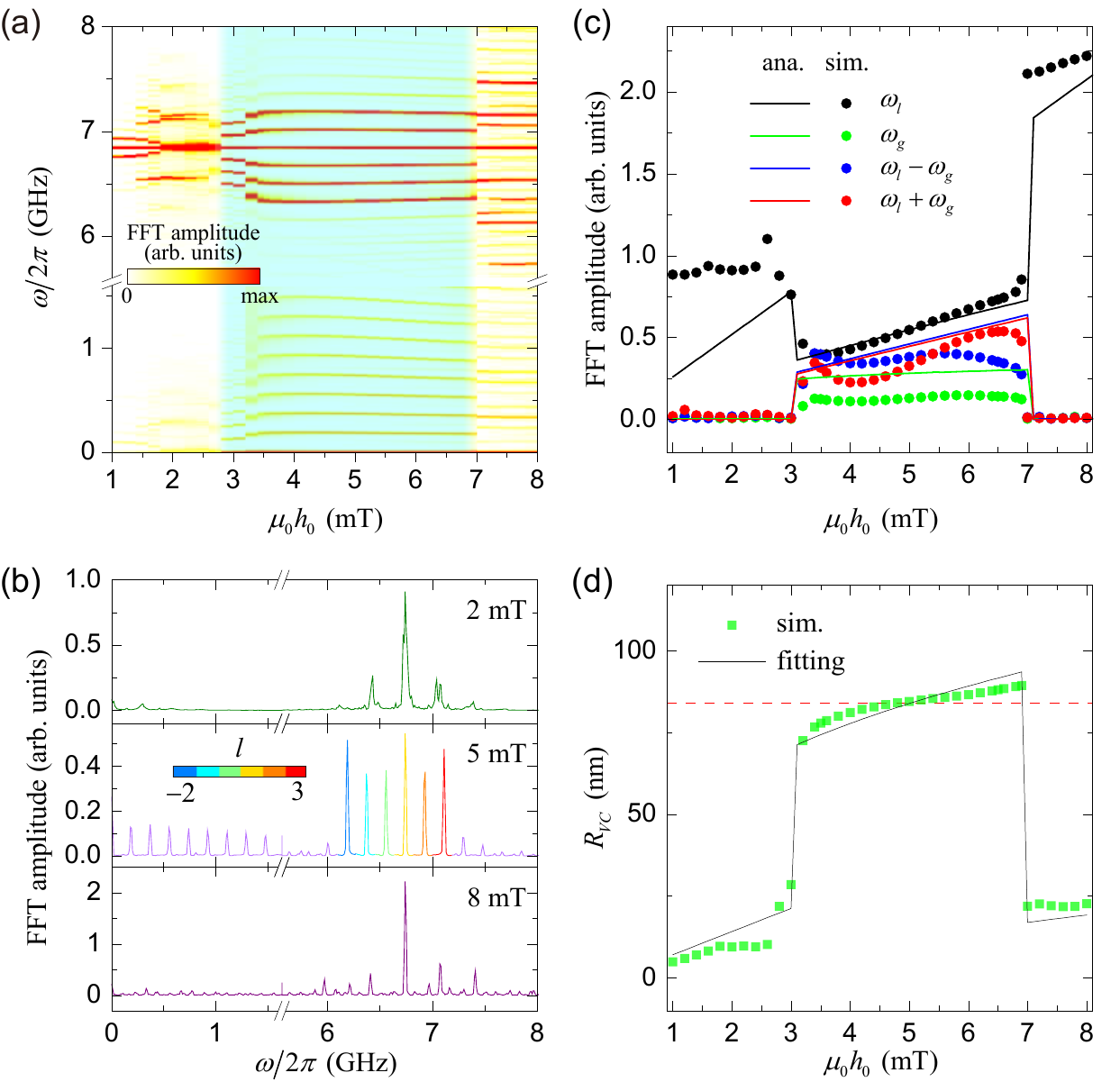}\\
  \caption{(a) Response of the vortex disk as a function of the driving-field amplitude $h_0$. The color codes the SW amplitude. The light cyan region denotes the emerging tMFC. (b) FFT spectra of the vortex disk under $\mu_0 h_0=2$, 5, and 8 mT, respectively. For $\mu_{0}h_{0}=$5 mT, a colorbar is used to label the OAM number carried by tMFC lines. (c) The amplitudes of four main peaks. Symbols are simulation results and curves are theoretical calculations. (d) VC gyration radius as a function of the driving-field amplitude $h_0$. The black curve denotes the analytical formula (\ref{eq_Rvc}). The red dashed line represents the orbital radius of the maximum SW amplitude.}\label{fig2}
\end{figure}

In this Letter, we study the quantization effect of the nonlinear magnon-vortex interaction in ferromagnetic nanodisks. We show that the circular geometry twists the magnon wave-function with spiral phase structure which acquires quantized OAM. Meanwhile, the there-magnon confluence and splitting scattering \cite{Aristov2016,Zhang2018} of twisted magnons off the gyrating VC generates a set of discrete and equally spaced spectral lines. We dub it twisted magnon frequency comb (tMFC). By applying a magnetic field perpendicular to the plane of the magnet, we discover an emerging magnonic Penrose superradiance when the disk thickness is larger than a critical value. In such a case, the higher-order modes of the tMFC are significantly amplified while the lower-order ones are trapped within the VC gyrating orbit manifesting as the ergoregion, which is analogous to the amplification of waves scattered by a rotating black hole.

\emph{Model}.---We start with the following Hamiltonian to model the vortex state in the nanodisk
\begin{equation}\label{eq_Hamiltonian}
  \mathcal{H}=\int\Big[A(\nabla \mathbf{m})^{2}-\frac{1}{2}\mu_{0}M_{s}\mathbf{m}\cdot\mathbf{H}_{d}-\mu_{0}M_{s}\mathbf{m}\cdot\mathbf{H}_{z}\Big]d\mathbf{r},
\end{equation}
where $\mathbf{m}=\mathbf{M}/M_{s}$ is the unit magnetization vector with the saturated magnetization $M_{s}$, $A$ is the exchange constant, $\mu_{0}$ is the vacuum permeability, $\mathbf{H}_{d}$ is the dipolar field, and $\mathbf{H}_{z}$ is the external bias magnetic field. Without applying $\mathbf{H}_{z}$, the competition between the exchange and dipolar interactions gives rise to magnetic vortex as a ground state in the nanodisk \cite{Ha2003}. To study the spin-wave dynamics in magnetic vortex, we assume a small fluctuation $\delta\mathbf{m}$ around the ground state. By performing the Holstein-Primakoff transformation, we can express the fluctuation as the bosonic operators $a^{\dagger}(\mathbf{r},t)$ and $a(\mathbf{r},t)$ corresponding to the magnon creation and annihilation, respectively. We focus on the third-order terms of boson operators that represent the three-magnon processes (see Supplemental Material \cite{Suppl}). If a twisted spin-wave mode ($\omega_{l}$) is resonantly excited by a strong microwave field, the gyrotropic mode ($\omega_{g}$) of the VC can be generated non-resonantly due to the mode splitting. It is noted that the gyrating VC carries an unit OAM quantum \cite{Mruczkiewicz2017}. Subsequently, the gyrotropic mode can hybridize with the twisted spin-wave mode through the three-magnon confluence process and generate the sum-frequency mode $\omega_{l}+\omega_{g}$ with OAM quantum number $l+1$. Similarly, the difference-frequency mode is $\omega_{l}-\omega_{g}$ with OAM quantum number $l-1$. This is the selection rule for nonlinear magnon-VC scatterings \cite{Suppl}. Further hybridizations generate higher-order spin-wave modes and finally lead to the formation of the tMFC with the mode spacing equal to the gyration frequency of the VC and the OAM difference by one between consecutive spectral lines, as illustrated in Fig. \ref{fig1}(a).

\emph{tMFC}.---To verify the above tMFC picture, we perform full micromagnetic simulations using MUMAX3 \cite{Vansteenkiste2014}. We consider a circular ferromagnetic disk of diameter $2R=300$ nm and thickness $d=5$ nm (if not stated otherwise), as indicated in Fig. \ref{fig1}(a).
Magnetic parameters of permalloy are used in simulations \cite{Suppl}.
To characterize the intrinsic spectrum of the vortex, we applied a sinc-function field $\mathbf{h}(t)=h_{0}\mathrm{sinc}(\omega_{c}t)\hat{x}$ with amplitude $\mu_0h_{0}=10$ mT and cutoff frequency $\omega_{c}/2\pi=20$ GHz. Results obtained from the fast Fourier transform (FFT) of $\delta m_{z}$ averaged over the whole disk are plotted in Fig. \ref{fig1}(b). Three main peaks are found at 0.18, 6.11, and 6.74 GHz, which correspond to the gyrotropic, and twisted spin-wave ($l=\mp1$) modes, respectively. A weak frequency-doubling effect of the gyrotropic mode is also observed at 0.36 GHz.

We then apply an in-plane rotating field $\mathbf{h}(t)=[h_{0}\cos(\omega_{0}t),h_{0}\sin(\omega_{0}t),0]$ with $\omega_{0}/2\pi=6.74$ GHz to excite the twisted spin-wave for the $l=+1$ mode. We make FFT of the time-dependent magnetization and obtain excitation spectra at various field amplitudes $h_0$, as shown in Fig. \ref{fig2}(a). Three main phases are identified: (i) Below 2.6 mT, only the driving mode and two side peaks are excited. The frequency spacing is not equal to 0.18 GHz ($\omega_g$), but 0.09 GHz ($0.5\omega_g$) below 1.2 mT and 0.3 GHz for field from 1.4 to 2.6 mT. We attribute it to non-resonant gyration of VC which carries a finite mass originating from its interaction with twisted magnons \cite{Suppl}. (ii) In the field window from 2.8 to 6.8 mT, a clear frequency comb emerges. The mode spacing, equal to the gyrotropic frequency, increases from 0.12 GHz at 2.8 mT to 0.18 GHz at 3.4 mT, then keeps 0.18 GHz up to 6.8 mT. (iii) Above 7 mT, the VC gyrates at 0.34 GHz ($\approx2\omega_g$) and the frequency comb disappears and is replaced by a main peak at 6.74 GHz along with chaotic side peaks [see Fig. \ref{fig2}(b)]. Figure \ref{fig2}(c) shows the amplitude of four major modes at 6.74 GHz (black dots), 0.18 GHz (green dots), 6.56 GHz (blue dots), and 6.92 GHz (red dots), respectively. One can observe the emergence of the gyrotropic ($\omega_{g}$), sum-frequency ($\omega_{l}+\omega_{g}$), and difference-frequency ($\omega_{l}-\omega_{g}$) modes between 3 and 7 mT, which is accompanied by the consumption of the injected $l=+1$ mode. Theoretical results [solid curves in Fig. \ref{fig2}(c)] based on the spin-wave dynamics counting three-magnon interactions \cite{Suppl} agree well with micromagnetic simulations.

We plot the VC gyration radius ($R_{\mathrm{VC}}$) as a function of the driving field amplitude in Fig. \ref{fig2}(d). It shows that $R_{\mathrm{VC}}$ has a large value ($\sim80$ nm) from 3 to 7 mT, which perfectly matches the field range of the MFC generation in Fig. \ref{fig2}(c).
It suggests that twisted magnon-vortex locking is essential for the generation of tMFC, i.e., the VC should gyrate along the orbits where the amplitude of twisted magnons is maximized [labeled by the red dashed line in Fig. \ref{fig2}(d)].
To analytically interpret the field-dependence of $R_{VC}$, we consider a generalized Thiele's equation
\begin{equation}\label{eq_Thiele}
  -\mathbf{G}\times\mathbf{\dot{R}}_{VC}-\mathcal{\hat{D}}\mathbf{\dot{R}}_{VC}+\frac{\partial W}{\partial\mathbf{R}_{VC}}=0,
\end{equation}
where $\mathbf{G}=G\hat{z}=-2\pi pqdM_{s}/\gamma\hat{z}$ is the gyrovector with the vortex polarity $p$, chirality $q$, and gyromagnetic ratio $\gamma$, $\mathcal{\hat{D}}$ is the damping tensor, and $\mathbf{R}_{VC}$ is the VC position. The potential $W(\mathbf{R}_{VC})=W(0)+\kappa/2|\mathbf{R}_{VC}|^{2}+\beta/4|\mathbf{R}_{VC}|^{4}-\mu(\hat{z}\times\mathbf{h}_{ext})\cdot\mathbf{R}_{VC}$, where $\kappa=10 \mu_{0}M_{s}^{2}d^{2}/9R$ is the stiffness coefficient, $\mu=2\pi RdM_{s}q/3$, $\mathbf{h}_{ext}$ is the external microwave field, and $\beta$ is the coefficient of the quartic parameter modeling the potential anharmonicity \cite{Suppl}. Substituting $W(\mathbf{R}_{VC})$ into Eq. (\ref{eq_Thiele}), we have
\begin{equation}\label{eq_Thiele2}
  -G\hat{z}\times\mathbf{\dot{R}}_{VC}-\mathcal{\hat{D}}\mathbf{\dot{R}}_{VC}+\kappa\mathbf{R}_{VC}+\beta|\mathbf{R}_{VC}|^{2}\mathbf{R}_{VC}-\mu(\hat{z}\times\mathbf{h}_{ext})=0.
\end{equation}
For a counterclockwise rotating microwave field $\mathbf{h}_{ext}=[h_{0}\cos(\omega_{0}t),h_{0}\sin(\omega_{0}t),0]$, Eq. (\ref{eq_Thiele2}) can be numerically solved \cite{Suppl}. It is found that the VC can be efficiently driven to a large gyration orbit only when the driving frequency is close to $\omega_{g}$. For $\omega_0=\omega_g$, the analytical expression of the steady-state gyration radius can be derived as
\begin{equation}\label{eq_Rvc}
  R_{VC}=\sqrt{\big(-\frac{c}{2}+\sqrt{\Delta}\big)^{\frac{1}{3}}+\big(-\frac{c}{2}-\sqrt{\Delta}\big)^{\frac{1}{3}}},
\end{equation}
where $c=-\mu^{2}h_0^2/\beta^2$ and $\Delta=(c/2)^2+(b/3)^3$ with $b=D^2\omega_g^2/\beta^2$.
This solution, however, is not fully consistent with simulation results [Fig. \ref{fig2}(d)], because the nonlinear interaction between the VC and twisted magnons was missed by Eq. (\ref{eq_Thiele2}). To accurately model the VC gyration driven by a non-resonant field, we treat the three-magnon ($g_q a_l a_q^\dagger + g_p a_l^\dagger a_p$) term in Eq. (S3b) \cite{Suppl} as an effective field oscillating with the frequency $\omega_\mathrm{eff}$ and amplitude $c_h h_0$. Here $c_h$ is the parameter describing the ratio of the effective field to driving field and the effective frequency is $\omega_g/2$, $\omega_g$, and $2\omega_g$ for the regime below, within, and above the tMFC, respectively. The VC gyration radius then can be well fitted by Eq. (\ref{eq_Rvc}) through a substitution $h_0\rightarrow c_h h_0$ with $c_h=0.31$ for the field range 3$\sim$7 mT and by numerically solving Eq. (\ref{eq_Thiele2}) with $c_h=0.015$ outside this regime [see the black curve in Fig. \ref{fig2}(d)].

\begin{figure}
  \centering
  \includegraphics[width=0.48\textwidth]{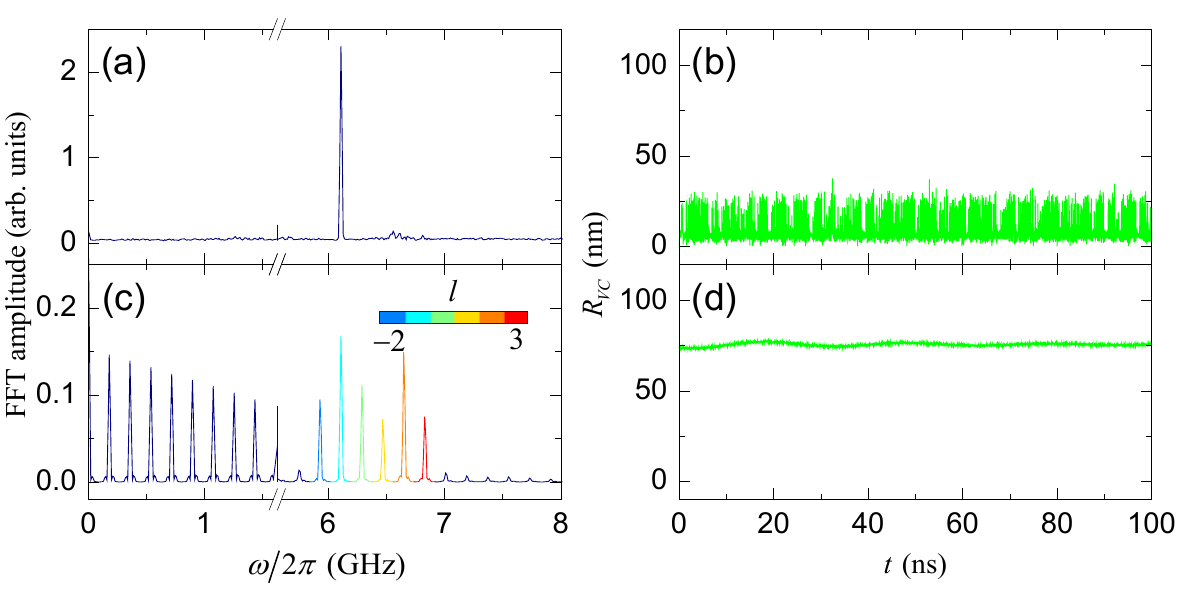}\\
  \caption{FFT spectra of the vortex disk under a single-frequency (a) and a dual-frequency (c) clockwise-rotating field. (b) and (d) are the corresponding VC gyrating radius in (a) and (c).}\label{fig3}
\end{figure}

We also study the spectra of the vortex disk driven by a clockwise rotating field ($\mathbf{h}=h_{0}[\cos(\omega_{0}t),-\sin(\omega_{0}t),0]$ with $\mu_0 h_0=10$ mT and $\omega_0/2\pi=6.11$ GHz corresponding to $l=-1$ mode). The tMFC around 6.11 GHz is not observed and the VC almost stays at the disk center without any gyration [see Figs. \ref{fig3}(a) and \ref{fig3}(b)]. This is because the natural VC gyration is counterclockwise and hardly driven by clockwise twisted magnons. However, by applying a dual-frequency rotating field to excite both the gyrotropic and $l=-1$ modes \cite{Suppl}, the tMFC is successfully generated, which highlights the importance of the VC gyration in the tMFC generation [Figs. \ref{fig3}(c) and \ref{fig3}(d)]. The comb structure is quite robust against thermal fluctuations \cite{Suppl}.

\begin{figure}
  \centering
  \includegraphics[width=0.49\textwidth]{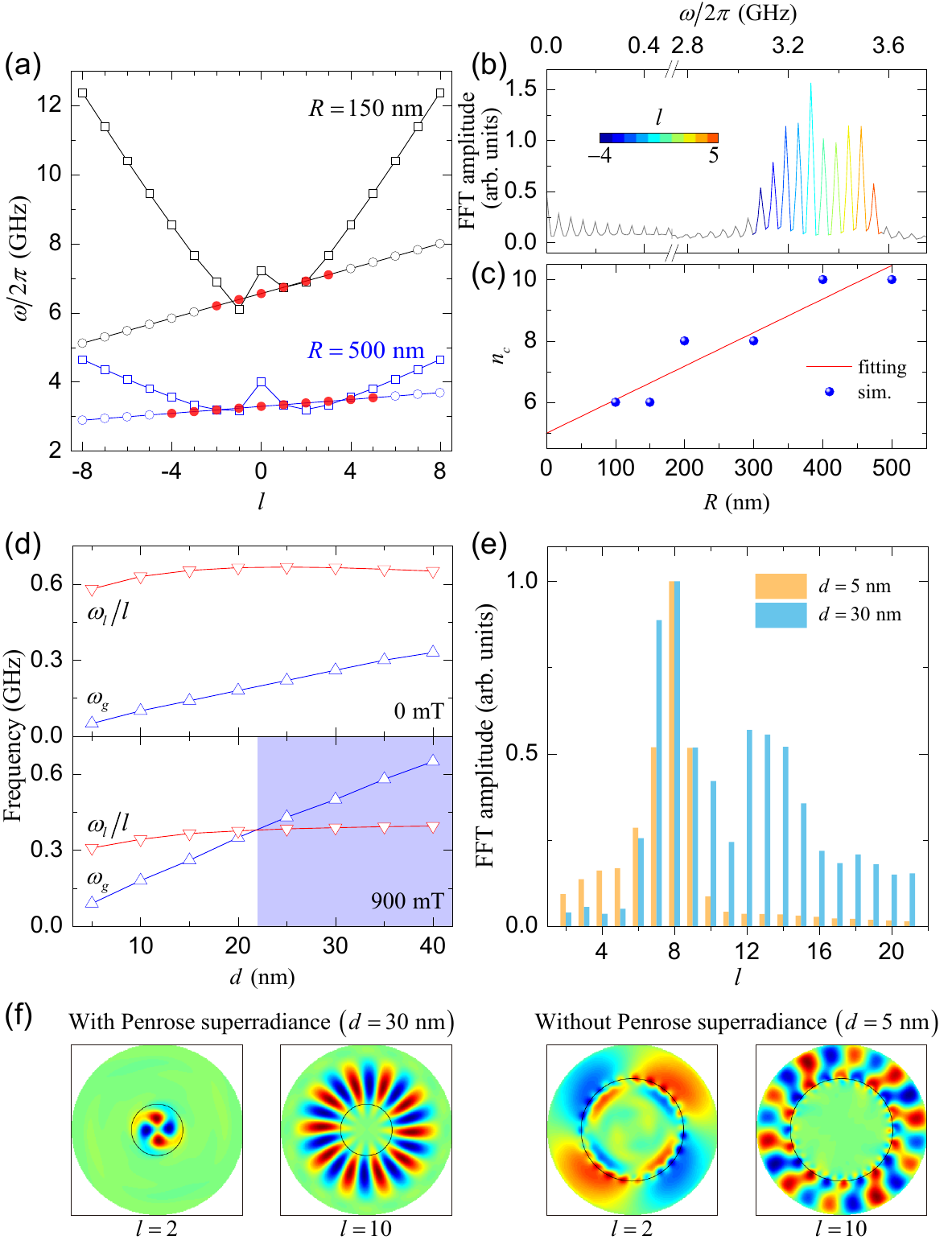}\\
  \caption{(a) The SW dispersion relation of the vortex disk with two radiuses. (b) FFT spectrum of the vortex disk with $R=500$ nm excited by a dual-frequency microwave. (c) The number of tMFC lines as a function of the disk radius. Blue dots are simulation results and the red line is the linear fitting. (d) The eigenfrequency of twisted magnon with $l=8$ and VC gyrotropic mode as a function of the disk thickness for two different bias fields. Condition (\ref{eq_Penrose}) is indicated by the light blue region. (e) Normalized amplitudes of tMFC spectral lines under the bias field 900 mT. (f) Spatial distributions of magnons in the tMFC with and without Penrose superradiance. The black circles labeling the VC gyrating orbit indicate the ergoregion.}\label{fig4}
\end{figure}

We note that only a few tMFC lines are generated in simulations. To obtain a coherent understanding, we plot the dispersion relation of the vortex disk in Fig. \ref{fig4}(a) (black squares). Theoretical (factual) tMFC modes are denoted by open circles (red dots). One can see that the frequencies of the generated tMFC modes (from $l=-2$ to $l=3$) are very close to the internal spectra of the vortex disk. Outside this region ($l<-2$ and $l>3$), the selection rules come into play: tMFC modes cannot be generated due to the strong frequency mismatch between tMFC modes and eigenmodes allowed by magnetic vortex.

The limited number of spectral lines in the tMFC  raises a critical challenge to the generation of twisted magnons with large OAM, which can be improved in ferromagnetic disks of a large size. We then simulate the tMFC in nanodisks with radius $R=500$ nm. Due to the weak coupling between the VC and twisted magnons in such a large disk, it is difficult to drive the VC gyration solely by the $l=+1$ twisted magnons. We thus apply a dual-frequency rotating field to excite both the gyrotropic VC mode and $l=+1$ magnon mode \cite{Suppl}. Their nonlinear mixing generates the tMFC [see Fig. \ref{fig4}(b)]. Compared with the case of $R=150$ nm, the number of the tMFC lines now doubles owing to the attraction of the tMFC and internal mode dispersions [see the blue squares and circles in Fig. \ref{fig4}(a)]. The OAM quantum number of tMFC varies from $l=-4$ to $l=+5$.
By systematically investigating the number of tMFC lines ($n_c$) for different radiuses ($R$), we find that the dependence of $n_c$ on $R$ can be linearly fitted by $n_c=c_0 R+n_0$ with $c_0=0.01$ and $n_0=5$ [shown in Fig. \ref{fig4}(c)].
Following this trend, we expect the generation of twisted magnons with higher OAM (e.g., $|l|\sim100$) in larger disks (e.g., $R\sim10~\mu$m) \cite{Suppl}.

\emph{Penrose superradiance}.---From Fig. \ref{fig4}(b), we observe that the strength of the spectral line drastically decreases with its order. However, a flat enough MFC is indispensable for magnonic high-precision metrology, microwave-to-optical frequency conversion, and many other promising applications. Here, we propose a Penrose superradiance mechanism to amplify twisted spin waves. Originally, Penrose superradiance represents a process that particles scattered from a rotating black hole extract energy at the expense of the black hole rotational energy \cite{Penrose1969}. The concept was later extended by Zel'dovich to the amplification of electromagnetic waves reflected from a rotating, absorbing metallic cylinder \cite{Zeldovich1971}. Recent experiments based on acoustic waves \cite{Cromb2020} and photon superfluids \cite{Braidotti2022} have observed this wave amplification phenomenon. To realize the magnonic Penrose superradiance, i.e., the amplification of twisted magnons, the following condition must be met
\begin{equation}\label{eq_Penrose}
\omega_g>\frac{\omega_l}{l}.
\end{equation}
This condition can be derived from the dynamical susceptibility of the vortex disk in a rotating frame \cite{Guslienko2006,Faccio2017}, written as
\begin{equation}\label{eq_susceptibility}
  \chi(\omega_l)=\chi(0)\frac{\omega_{g}}{\omega_{g}-\omega^{\prime}-id_{\alpha}\omega^{\prime}},
\end{equation}
where $\chi(0)=2\gamma M_s/9\omega_g$ is the static susceptibility, $d_{\alpha}=-D/G$ is the effective damping parameter, and $\omega^{\prime}=\omega_{l}-l\omega_g$ due to the rotational Doppler frequency shift. When the damping term ($d_{\alpha}\omega^{\prime}$) becomes negative, an energy gain appears to amplify the twisted magnons. This consideration naturally recovers the Penrose criteria Eq. (\ref{eq_Penrose}).

However, we note that the eigenfrequency of twisted magnons usually is one order of magnitude higher than that of the VC gyration. As we proceed to realize the Penrose condition Eq. (\ref{eq_Penrose}), it is found that the gap between the VC gyration and twisted magnons decreases with the increase of the disk size ($R$ and $d$) and OAM number ($l$) and can be further reduced by applying an external bias magnetic field [see Fig. \ref{fig4}(d)]. We here choose the driving mode with a relatively large OAM ($l=8$) and apply a perpendicular bias magnetic field ($H_z=900$ mT) in a thick ferromagnetic disk ($d=$ 30 nm) of a cone vortex state \cite{Ivanov2002}.
The tMFC indeed becomes much more flat due to the amplification of higher-order modes [compare the cyan and orange histograms in Fig. \ref{fig4}(e)].
Intuitively, the gyrating orbit of VC acts like a energy barrier that resembles as an effective ergoregion. The lower-order modes lose energy and are trapped within the ergoregion, while the higher-order modes gain energy from the gyrating VC, thereby overcome the barrier to escape, see the left panel of Fig. \ref{fig4}(f).
For the case without the Penrose superradiance (e.g., $d=$ 5 nm), such a wave trapping phenomenon is absent, see the right panel of  Fig. \ref{fig4}(f) \cite{Suppl}.

\emph{Conclusion}.---In summary, we have investigated the quantization effect of the nonlinear three-magnon interactions between the gyrotropic VC and twisted magnons in ferromagnetic nanodisks. We predicted an emerging tMFC due to a strong overlap between VC gyrating orbit and twisted magnon wave-function maxima. The selection rule governing the confluence and splitting processes in circular geometries was derived. It was found that the frequency spacing in the tMFC is equal to the gyrating frequency of the VC and the OAM quantum number differs by one for neighboring spectral lines. We proposed the magnonic Penrose superradiance to amplify the higher-order twisted magnons and to trap the lower-order ones inside the ergoregion, i.e., the VC gyrating orbit. Our results suggest a promising approach to generate twisted magnons with very high OAM and to significantly improve the flatness of the MFC. This work also opens the door to the study of the fundamental physics of rotating black holes in magnonic platforms \cite{Yuan2022}.

\begin{acknowledgments}
This work was funded by the National Natural Science Foundation of China (Grants No. 12074057, No. 11604041, and No. 11704060). Z.W. acknowledges the financial support from the China Postdoctoral Science Foundation under Grant No. 2019M653063. H.Y.Y. acknowledges the European Union's Horizon 2020 research and innovation programme under Marie Sk{\l}odowska-Curie Grant Agreement SPINCAT No. 101018193.
\end{acknowledgments}

\end{document}